\documentstyle[aps,manuscript,12pt,graphics]{revtex}

\makeatletter

\providecommand{\LyX}{L\kern-.1667em\lower.25em\hbox{Y}\kern-.125emX\@}

\makeatother

\begin{document}

\title{Low-temperature orientational order and possible domain structures in C\( _{60} \)
fullerite }

\author{Vadim M. Loktev,\( ^{1,2} \)\thanks{
Author to whom correspondence should be addressed
} Yuri G. Pogorelov,\( ^{2} \) and Julia N. Khalak\( ^{1} \)}

\address{\( ^{1} \)Bogolyubov Institute fot Theoretical Physics, National Academy of
Sciences of Ukraine, Metrologichna str. 14b, Kiev 143, 03143 Ukraine; \( ^{2} \)Centro
de Física do Porto, Universidade do Porto, Rua do Campo Alegre 687, 4169-007
Porto, Portugal.}

\maketitle

\begin{abstract}
Based on a simple model for ordering of hexagons on square planar lattice, an
attempt has been made to consider possible structure of C\( _{60} \) fullerite
in its low temperature phase. It is shown that hexagons, imitating fullerens
oriented along \( C_{3} \) axes of \emph{sc} lattice, can be ordered into an
ideal structure with four non-equivalent molecules in unit cell. Then the energy
degeneracy for each hexagon rotations by \( \pi /3 \) around its \( C_{3} \)
axis leaves the translational and orientational order in this structure, but
leads to a random distribution of \( \pi /3 \) rotations and hence to {}``averaged{}''
unit cell with two molecules. However the most relevant structural defects are
not these intrinsic \char`\"{}misorientations\char`\"{} but certain walls between
the domains with different sequencies of the above-mentioned two (non-ideal)
sublattices. Numeric estimates have been made for the anisotropic inter-molecular
potential showing that the anisotropy is noticeably smaller for molecules in
walls than in domains.
\end{abstract}

\section{Introduction}

Study of equilibrium thermodynamic properties of C\( _{60} \) fullerite remains
among actual topics in the low temperature physics. In particular, the recent
experiments on its heat conduction \cite{1} and linear thermal expansion \cite{2}
revealed the anomalies proper just to this unique object. This relates to the
following properties observed in experiments.

i) Short enough (\( \sim  \)50 intermolecular spacings) mean free path for
acoustic phonons evidencing presence of a rather high amount (up to 10\%) of
structural or impurity scatterers, despite only less then \( 10^{-2} \) wt.\%
impurities are present in the initial material, 

ii) Negative (and really huge, up to \( 10^{2} \) ) value of Grüneisen coefficient
in solid C\( _{60} \) at \( T\sim 10 \)K.

In particular, to explain the low temperature behavior of heat conductance in
nominally pure C\( _{60} \) fullerite, scattering processes of phonon heat
carriers by some defects of orientational nature were invoked in Ref. \cite{1}.
Namely, it was supposed that with cooling the crystal some single, \char`\"{}orientationally
disordered\char`\"{} C\( _{60} \) molecules leave quenched in it. Then their
relative number should reach several percents, or in other words, so many molecules
become \char`\"{}orientational impurities\char`\"{} that one of them can be
found among nearest neighbors of each \char`\"{}regular\char`\"{} molecule.
This was justified by the estimates showing that if anisotropic part of inter-molecular
interaction (AIMI) contains two minima with relatively small energy difference,
but separated by a high enough (\( \approx 3\cdot 10^{3} \)K) energy barrier,
a considerable number of molecules can left frozen in the metastable state at
\( T \)\( \sim  \) \( 10^{2} \) K and a reasonable cooling rate. However,
estimates based on single-particle treatment considering relaxation of each
molecule independent of others, at fixed (static) environment, hardly can be
consistent. All the molecules are equivalent in the crystal, equally and self-consistently
participating in formation of the crystalline (molecular) field on each of them,
therefore the barriers should also depend on the relaxing molecules themselves.
Subsequently, energy estimates for several particular orientations of a single
molecule \cite{3} can hardly give a proper value of shortest time of escape
from its metastable state.\footnote{%
The collective character of crystalline modes should be also taken into account.
A well-known example is vibrational or magnetic spectra resulting from single
particle levels with excitation gap of the order of inter-particle interaction,
which is considerably softened (down to Goldstone gapless behavior) after collectivization.
Orientational modes are not exception in this sense.
}At least, it should be noted that a great number of misoriented molecules can
transform the crystal into an \char`\"{}orientational solution\char`\"{} or
even into a glass (if this will be accompanied by unlimited extension of relaxation
times spectrum). That idea of orientational glass and resulting competition
between isotropic and anisotropic parts of intermolecular potential was proposed
in Ref. \cite{4} to explain the anomalous large negative thermal expansion
of solid C\( _{60} \) discovered by Aleksandrovskii et al. \cite{2}. However,
at present no numeric estimates are available for this mechanism that could
confirm the observed expansion.

The above mentioned problems with justification of the proposed physics of anomalous
thermal behavior of solid C\( _{60} \) suggest one to seek some alternatives,
more compatible with the translational invariance of a crystal. In this communication,
one such mechanism is proposed, related to possible existence of several orientational
domains in \emph{sc} phase of solid C\( _{60} \), separated by well defined
domain walls. The latter could play the role of effective scatterers for phonon
heat carriers. Besides, the higher symmetry of local crystalline field on C\( _{60} \)
within the walls can restore the conditions for their almost free rotation,
which is necessary (see Ref. \cite{2} and references therein) to account for
negative thermal expansion.

\section{Model}

It is well known (see, for example the reviews \cite{5},\cite{6}) that below
the point of orientational melting \( T_{m}^{(high)}\approx 260 \) K the \emph{fcc}
lattice of C\( _{60} \) fullerite is divided into four \emph{sc} sublattices
with one of the molecule \( C_{3} \) axes oriented along one of the cube diagonals
(which also are crystalline \( C_{3} \) axes). It is of interest to notice
that the corresponding \( Pa3 \) structure, characteristic for simplest molecular
cryocrystals where small quadrupole-quadrupole interactions dominate \cite{7},
permits to assume presence of an induced quadrupole moment on C\( _{60} \)
in fullerite, despite its complete absence for free C\( _{60} \) molecule.\footnote{%
Evidently, a number of thermal rotational excitations of C\( _{60} \) molecules
are present at \( T>T_{m}^{(high)} \), all of them associated with certain
multipole distortions. The lowest energies (\( \sim 10^{2} \) cm\( ^{-1} \)
\cite{8}) relate to intra-molecular quadrupole vibrations, indistinguishable
from rotations. Self-consistent admixture of these excitations into the molecule
ground state in the crystal field reduces (below \( T_{m}^{(high)} \)) the
almost spherical symmetry down to an axial, contributing to the total energy
gain. So one can suppose that below \( T_{m}^{(high)} \) the definite orientation
of fullerens along a \( C_{3} \) axis is fixed and long-ranged, corresponding
to the common order of quadrupoles. Otherwise, the \emph{sc} lattice cannot
be realized.
}Also it can be expected that no \char`\"{}transversal\char`\"{} order with respect
to each of these axes takes place until the low temperature transition at \( T_{m}^{(low)}\approx 90 \)
K. But since the molecules C\( _{60} \) present truncated icosahedra, having
5th order axes among their symmetry elements, they cannot be completely ordered
into \emph{sc} lattice by the impossibility of simultaneous optimization of
local (crystal-field) and inter-molecular potentials. Therefore certain kinds
of defects are inevitable at low temperatures, either point (individual) or
extended (collective).

The first type of defects is usually related to some local disturbance of structural
or compositional order, while the second (as dislocations, domain walls, twin
boundaries, etc.) can exist even at fully uniform background. Local disorder
in fullerite could be due to, for example, isotopically substituted C\( _{60} \)
molecules, or C\( _{n} \) fullerenes with \( n\neq 60 \), or impurities like
H\( _{2} \). But the samples of C\( _{60} \) fullerite with the above-mentioned
anomalies of low temperature properties were especially prepared and purified
so that there were no physical reasons for any noticeable contents of local
defects (including misoriented molecules).

Then a more plausible source of low temperature anomalies can be sought in extended
(topological) defects, and in view of the possibility of several energetically
equivalent domain structures to exist under reduced cubic symmetry, these defects
can be associated with the walls between such domains.

Of course, even a simple cubic lattice made of so complex and symmetrical molecules
as C\( _{60} \) presents great technical difficulties for straightforward calculation
of full inter-molecular potential, defined by high-order multipole moments with
great number of components, and of related low-energy (non-linear) excitations
in the crystal. Hence, not pretending to give quantitative predictions for real
fullerite, we limit ourselves below to consideration of a simplified model including
the relevant features of fullerite: reduction of the crystalline point symmetry
by its incompatibility with the molecular symmetry, a double-well potential
of AIMI, and the related possibility for domains and domain walls.

Let us consider a system of flat hexagonal molecules (imitating C\( _{60} \)
molecules seen along \( C_{3} \) axis\footnote{%
Since the same 3rd order rotational symmetry holds for fullerene molecules being
projected at cube faces.
}) located in sites of rigid square planar (\emph{sp}) lattice, modeling 3D \emph{fcc}
lattice. To evaluate the angular part of pair interaction between electrically
neutral hexagons, we suppose two kinds of negative charges, \( -\left( 1\pm \alpha \right)  \),
located at centers of hexagon sides and unit positive charges at their vertices
(see Fig. 1a). Such distribution of negative charges recalls single covalent
bonds at borders between pentagon and hexagon and double bonds between two hexagon
rings in a truncated icosahedral molecule. Here the charge and geometric asymmetry
parameter \( \alpha  \), reducing the \( C_{6} \) symmetry of a hexagon down
to \( C_{3} \), reflects one of the most important features of real C\( _{60} \)
fullerene, \( 120 \)\( ^{\circ } \) alternation of such rings around each
its hexagon.

\begin{figure}
{\par\centering \includegraphics{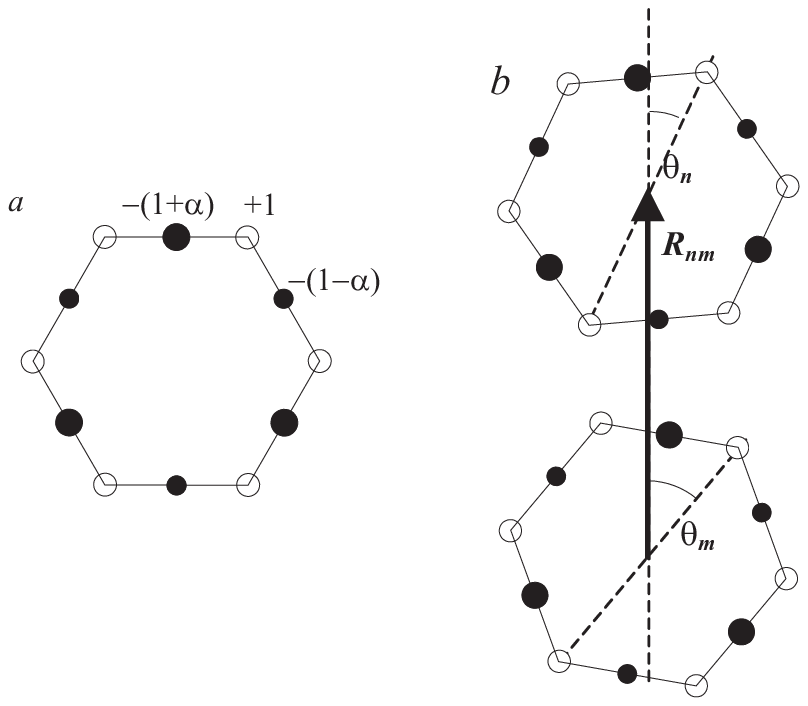} \par}

\caption{\label{fig1}Charge distribution adopted for the model hexagon molecule (\emph{a})
and possible orientations of two molecules with centers positioned at \protect\( \mathbf{n}\protect \)
and \protect\( \mathbf{m}\protect \) (\emph{b}).}
\end{figure}

The full Coulomb energy of a pair of hexagons (Fig. \ref{fig1}b) reads:\begin{equation}
\label{1}
V_{\mathbf{nm}}\left( \theta _{\mathbf{n}},\theta _{\mathbf{m}}\right) =\sum _{\mu ,\sigma }V_{\mathbf{nm}}^{\left( \mu \sigma \right) }\left( \theta _{\mathbf{n}},\theta _{\mathbf{m}}\right) ,
\end{equation}
where the indices \( \mu ,\sigma  \) take the values \( v \), \( b \), or
\( B \), related to vertices and to bonds with smaller and greater negative
charges, respectively, and the particular terms are:\[
V_{\mathbf{nm}}^{\left( vv\right) }\left( \theta _{\mathbf{n}},\theta _{\mathbf{m}}\right) =\sum _{j,k=0}^{5}\left\{ \left[ R_{\mathbf{nm}}+\cos \left( \theta _{\mathbf{n}}+\frac{\pi j}{3}\right) -\cos \left( \theta _{\mathbf{m}}+\frac{\pi k}{3}\right) \right] ^{2}+\right. \]
\begin{equation}
\label{2}
\left. +\left[ \sin \left( \theta _{\mathbf{n}}+\frac{\pi j}{3}\right) -\sin \left( \theta _{\mathbf{m}}+\frac{\pi k}{3}\right) \right] ^{2}\right\} ^{-1/2},
\end{equation}

\[
V_{\mathbf{nm}}^{\left( vb\right) }\left( \theta _{\mathbf{n}},\theta _{\mathbf{m}}\right) =-\left( 1-\alpha \right) \sum _{j=0}^{5}\sum _{k=0}^{2}\left\{ \left[ R_{\mathbf{nm}}+\cos \left( \theta _{\mathbf{n}}+\frac{\pi j}{3}\right) -\right. \right. \]
\[
\left. \left. \frac{\sqrt{3}}{2}\cos \left( \theta _{\mathbf{m}}+\pi \frac{4k+1}{6}\right) \right] ^{2}+\left[ \sin \left( \theta _{\mathbf{n}}+\frac{\pi j}{3}\right) -\frac{\sqrt{3}}{2}\sin \left( \theta _{\mathbf{m}}+\pi \frac{4k+1}{6}\right) \right] ^{2}\right\} ^{-1/2}=\]
\begin{equation}
\label{3}
=V_{\mathbf{mn}}^{\left( bv\right) }\left( \theta _{\mathbf{m}},\theta _{\mathbf{n}}\right) =\frac{1-\alpha }{1+\alpha }V_{\mathbf{nm}}^{\left( vB\right) }\left( \theta _{\mathbf{n}},-\theta _{\mathbf{m}}\right) =\frac{1-\alpha }{1+\alpha }V_{\mathbf{mn}}^{\left( Bv\right) }\left( \theta _{\mathbf{m}},-\theta _{\mathbf{n}}\right) ,
\end{equation}
\[
V_{\mathbf{nm}}^{\left( bb\right) }\left( \theta _{\mathbf{n}},\theta _{\mathbf{m}}\right) =\left( 1-\alpha \right) ^{2}\sum _{j,k=0}^{2}\left\{ R_{\mathbf{nm}}+\frac{\sqrt{3}}{2}\left[ \cos \left( \theta _{\mathbf{n}}+\pi \frac{4j+1}{6}\right) -\right. \right. \]
\[
\left. \left. \left. -\cos \left( \theta _{\mathbf{m}}+\pi \frac{4k+1}{6}\right) \right] \right] ^{2}+\frac{3}{4}\left[ \sin \left( \theta _{\mathbf{n}}+\pi \frac{4j+1}{6}\right) -\sin \left( \theta _{\mathbf{m}}+\pi \frac{4k+1}{6}\right) \right] ^{2}\right\} ^{-1/2}=\]
\begin{equation}
\label{4}
=\frac{1-\alpha }{1+\alpha }V_{\mathbf{nm}}^{\left( bB\right) }\left( \theta _{\mathbf{n}},\theta _{\mathbf{m}}\right) =\frac{1-\alpha }{1+\alpha }V_{\mathbf{nm}}^{\left( Bb\right) }\left( -\theta _{\mathbf{n}},-\theta _{\mathbf{m}}\right) =\left( \frac{1-\alpha }{1+\alpha }\right) ^{2}V_{\mathbf{nm}}^{\left( BB\right) }\left( -\theta _{\mathbf{n}},-\theta _{\mathbf{m}}\right) ,
\end{equation}
\( R_{\mathbf{nm}}=\left| \mathbf{n}-\mathbf{m}\right|  \) is the distance
between the centers of hexagons on the sites \( \mathbf{n} \) and \( \mathbf{m} \)
of \emph{sp} lattice; \( \theta _{\mathbf{n}},\theta _{\mathbf{m}} \) are the
relative orientation angles; the distance from center to vertex is unity. It
can be noticed that, due to \( C_{3} \) symmetry of charges in a hexagon, the
clockwise and counterclockwise rotations are not equivalent in AIMI.

Despite the simplified geometry of \emph{sp} lattice of hexagons and the neglecting
of quantum effects (charge delocalization, covalency, etc.), one can expect
this rather rough model to give a correct qualitative behavior of AIMI and its
dependence on the charge distribution within the molecule and a reasonable estimate
of contributions from different mutual configurations of molecules.

\section{Pair interactions and ordering types}

The numerical results for AIMI, Eq. \ref{1}, are shown in Fig. 2 for some typical
mutual configurations and several values of the asymmetry parameter \( \alpha  \).
First of all, it is seen that, for \( \alpha \neq 0 \), AIMI for two hexagons
possesses a distinct 120\( ^{\circ } \) periodicity and two-hump profiles.
This reflects correctly AIMI for two C\( _{60} \) molecules where a double-well
potential describes the so-called pentagon and hexagon configurations \cite{9}
(see also \cite{5},\cite{6}). It is also seen that with decreasing asymmetry
of negative charge distribution, AIMI becomes smoother, though some minima (see
Fig. 2a,b) become deeper, so that in the limit \( \alpha \rightarrow 0 \) all
the minimum energies are equal and negative.

\begin{figure}
{\par\centering \resizebox*{12.5cm}{19cm}{\includegraphics{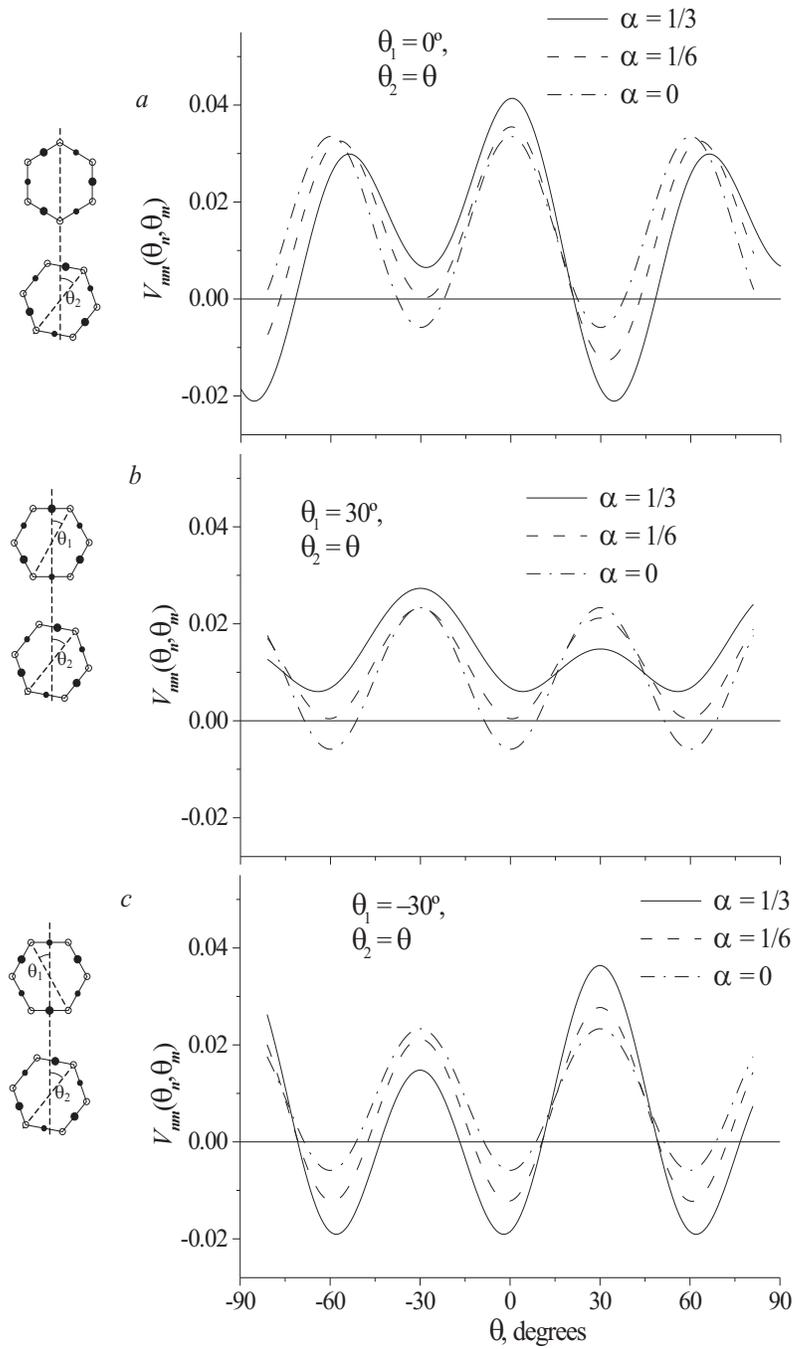}} \par}

\caption{\label{fig2}AIMI fo most characteristic (shown at left) mutual configurations
of hexagons at fixed orientation of one of them. The axes and rotation angles
correspond to Fig. \ref{fig1}b. The intermolecular distance \protect\( R_{12}\protect \)
was chosen 3 (in units of hexagon side).}
\end{figure}

It follows from Fig. \ref{fig2} that for all asymmetry values, except \( \alpha =0 \),
the most stable configuration is that where a vertex of one molecule points
to a greater negative charge of neighbor molecule (Fig. 2a,c) while the maximum
repulsion corresponds to parallel neighboring sides with such negative charges.
At least, in the case \( \alpha =0 \) the 60\( ^{\circ } \) periodicity corresponding
to \( C_{6} \) axis restored, nevertheless leaving the same (vertices against
sides) most stable configuration.

\begin{figure}
{\par\centering \includegraphics{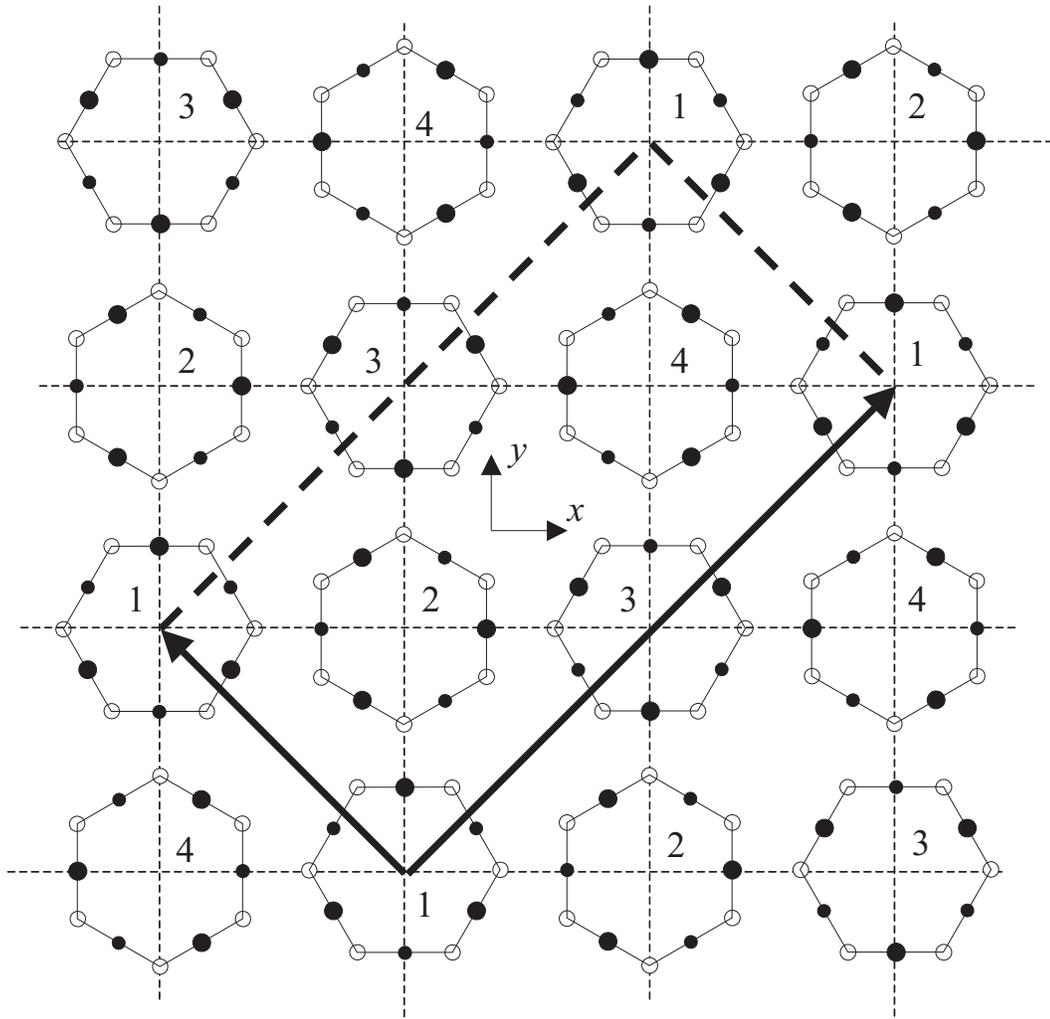} \par}

\caption{\label{fig3}An example of ordering of hexagons into the \emph{sp} lattice
with four molecules in unit cell and its translation vectors. Equivalent structures
can be obtained by all permutations preserving opposite parities between nearest
neighbors.}
\end{figure}

Knowledge of pair interaction and most stable configurations for two hexagons
enables one to order them in a \emph{sp} lattice. Then AIMI requires that one
of long axes of each hexagon be oriented along a crystalline axis and its nearest
neighbors be rotated by \( \pi /6 \). This readily divides the sp lattice into
two inter-twinned ones, with long hexagon axes aligned with \( x \) (\char`\"{}horizontal\char`\"{},
H) and \( y \) (\char`\"{}vertical\char`\"{}, V), respectively. But taking
into account that a molecule has two non-equivalent positions with respect to
negative charges for each alignment, the ideal order in \emph{sp} lattice of
such hexagons corresponds to \char`\"{}parquets\char`\"{} (one of them is shown
in Fig. \ref{fig3}) with four molecules in unit cell, two horizontal, denoted
1 and 3, and two vertical, 2 and 4. Then each of the two above-mentioned sublattices
contains only even or odd positions. Here the long-range order holds not only
for translations and orientations but also for the charge pattern. It should
be also noted that, because of incompatible point groups for asymmetric hexagons
and \emph{sp} lattice, it is impossible to arrange all nearest neighbors of
each hexagonal molecule in positions with maximum negative AIMI. Though some
its neighbors occur in metastable minima of AIMI, nevertheless the total energy
balance proves to be negative and stable.

\begin{figure}
{\par\centering \includegraphics{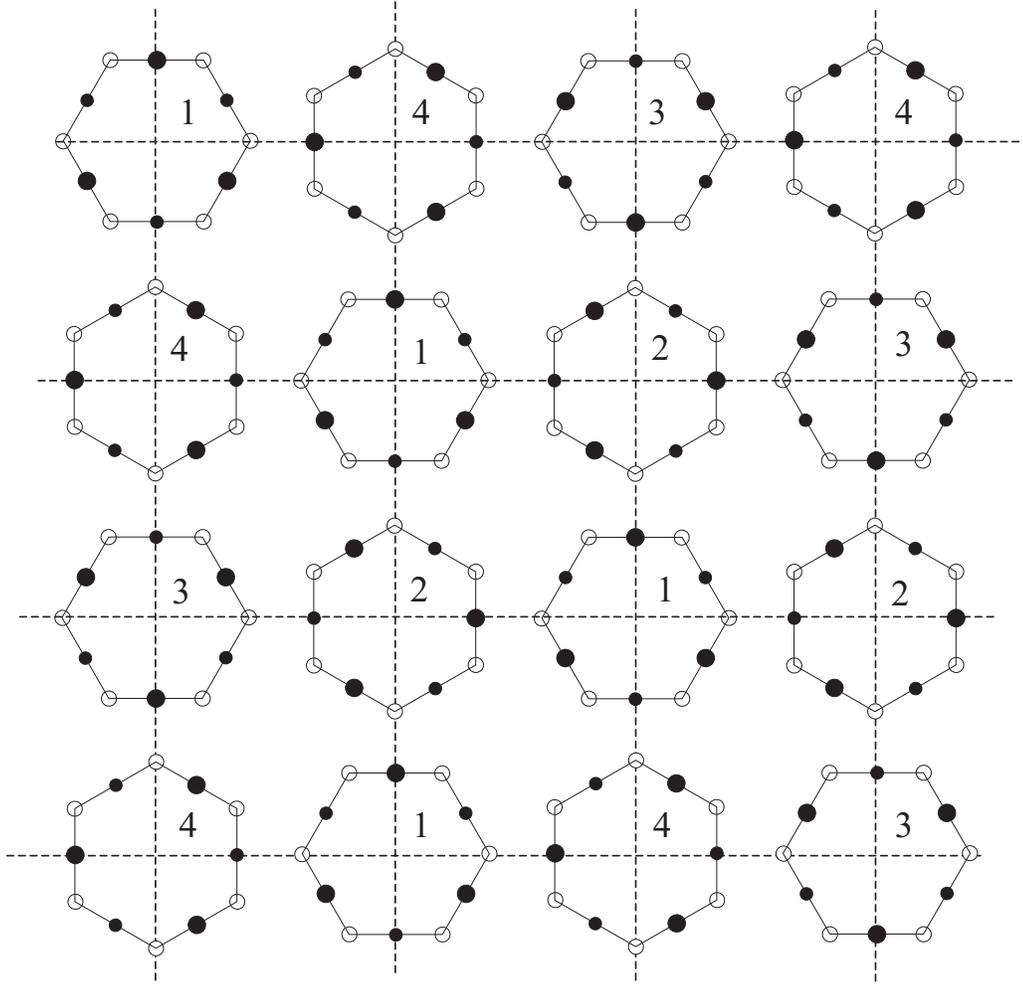} \par}

\caption{\label{fig4}Fragment of \emph{sp} lattice structure obtained by random substitutions
\protect\( 1\leftrightarrow 3\protect \) and \protect\( 2\leftrightarrow 4\protect \)
introduced into the ideal 4-sublattice structure in Fig. 3.}
\end{figure}

This kind of order is peculiar by its frustration, or the energy degeneracy
with respect to the substitutions \( 1\leftrightarrow 3 \) and \( 2\leftrightarrow 4 \).
These transformations are just generated by a \( C_{6} \) rotation, which is
not an element of symmetry group of a molecule with asymmetric charge distribution
(see Fig. \ref{fig1}a). In its turn, this implies that the \emph{sp} lattice
of hexagons, preserving the above described translational and orientational
order,\footnote{%
Strictly speaking, the \( C_{6} \)-rotated molecule can change its distances
to nearest neighbors, but we ignore this virtually small effect in view of average
translational invariance of the lattice. At the same time, the AIMI analysis
shows that the orientational order is not perturbed even under \( C_{6} \)
rotations.
}can be created in a thermodynamical way with a specific disorder left within
even and odd sites. This transforms the ideally ordered 4-sublattice into a
non-ideal 2-sublattice structure like a simulated fragment shown in Fig. \ref{fig4}.
In such a crystal the \( C_{6} \) rotation intrinsically enters the point symmetry
group of a molecule.

Evidently, 4- or 2-sublattice structures admit existence of several equivalent
arrangements with permutated sublattices separated by certain extended defects,
domain walls or anti-phase boundaries. These defects might effectively contribute
into low temperature thermal properties of the system. Below we consider an
example of such a defect in a 2-sublattice structure.

\section{Domain wall structure}

The above-indicated structure of 2-sublattice ordering of hexagons in sp lattice
provides equal conditions for all of them, and the mentioned disorder does not
result in any characteristic isolated defects. This is also seen from Fig. \ref{fig2}
showing rather high barriers between stable and metastable minima. Hence each
hexagon, either in 4- or 2- sublattice structure stays near an AIMI minimum
that defines its libration spectrum.

However this does not prevent defects at all in such a crystal. In particular,
in the course of thermodynamical growth, there can appear, as usual, some vacancies
and dislocations (which will not be discussed here) and also a specific kind
of defects, the anti-phase boundaries, characteristic for any multi-sublattice
orientational (vector or tensor) structure. They emerge between the regions,
identical in their coordination but different in the attribution of molecule
orientations to sublattices.

\begin{figure}
{\par\centering \resizebox*{10cm}{14cm}{\includegraphics{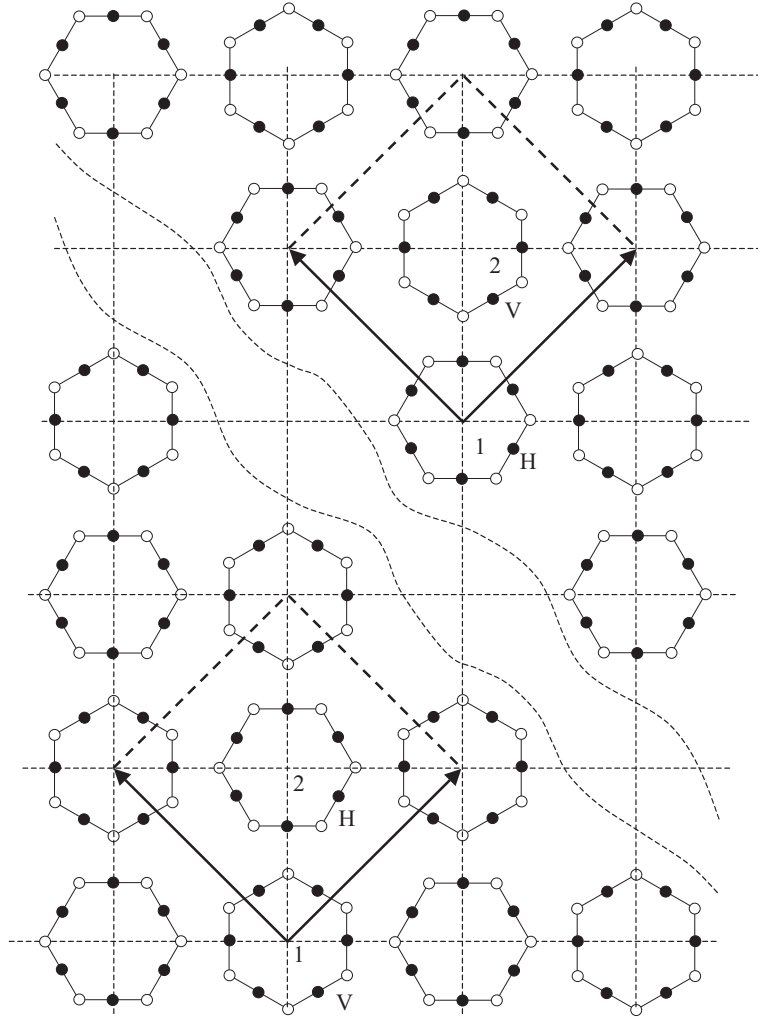}} \par}

\caption{\label{fig5}A schematic of growths and approaching of two domains with orientationally
ordered subsystems of hexagons. In the domain I (upper right) vertically oriented
molecules (V, \protect\( \theta =0\protect \)) are located in the sites of
the 1st sublattice and horizontal (H, \protect\( \theta =\pi /6\protect \))
in 2nd sublattice, and in the domain II (lower left) the attribution is inverse.}
\end{figure}

Actually, the transition from rotation to libration of molecules is a 1st kind
transition, realized through formation of nuclei (domains) of orientational
order with definite attribution of sublattices to molecule orientations. Extension
of such domains (see Fig. \ref{fig5}) leads them to contact each other, forming
a continuous ordered structure. There are two possible modes of such a \char`\"{}meeting\char`\"{}.
At the contacts \( \ldots  \)HVHV\( \rightarrow \leftarrow  \)HVHV\( \ldots  \)
or \( \ldots  \)VHVH\( \rightarrow \leftarrow  \)VHVH\( \ldots  \) the two
structures perfectly match, producing a single coherent domain. But the contacts
\( \ldots  \)VHVH\( \rightarrow \leftarrow  \)HVHV\( \ldots  \) or \( \ldots  \)HVHV\( \rightarrow \leftarrow  \)VHVH\( \ldots  \)
produce a mismatch, so that the closest molecules to the boundary should be
orientationally tuned to provide a continuous transition from one domain to
another. Evidently, far from the boundary such domains are indistinguishable
and the boundary itself is just a consequence of the initial conditions of the
growth. Moreover, no visible thermodynamical mechanisms for domain structure
formation (like those known, for instance, in 2-sublattice antiferromagnets
\cite{10},\cite{11}) can be indicated in this system of orientationally ordered
hexagons.

To describe consistently the 2-sublattice structure, let us redefine the orientation
angle \( \theta _{\mathbf{n},i} \) for an \char`\"{}averaged\char`\"{} molecule
(possessing \( C_{6} \) symmetry) at \( i \)th site in \( \mathbf{n} \)th
unit cell as the smallest positive angle between one of its vertices and \( y \)-axis
(see Fig. \ref{fig1}b). Then for each unit cell we can naturally define the
two angles\begin{equation}
\label{5}
\varphi _{\mathbf{n}}=\theta _{\mathbf{n},2}-\theta _{\mathbf{n},1},\qquad \psi _{\mathbf{n}}=\theta _{\mathbf{n},2}+\theta _{\mathbf{n},1},
\end{equation}
which play the role of order parameters. For the two fragments of ordered structures
shown in Fig. \ref{fig5} the corresponding values are uniform in space: \( \varphi _{\mathbf{n}}=\varphi _{I}=\pi /6 \),
\( \psi _{\mathbf{n}}=\psi _{I}=\pi /6 \) in the domain I, and \( \varphi _{\mathbf{n}}=\varphi _{II}=-\pi /6 \),
\( \psi _{\mathbf{n}}=\psi _{II}=\pi /6 \) in the domain II. Thus the two domains
are distinguished by the inversion of parameter \( \varphi  \), like 180\( ^{\circ } \)
domains in a 2-sublattice antiferromagnet. 

One can built a boundary between these two domains, located at the origin and
characterized by the unit normal vector \( \mathbf{d} \), so that \( \varphi _{\mathbf{n}} \)
changes when \( \mathbf{n} \) crosses the domain wall, reaching asymptotic
values \( \varphi _{\mathbf{n}}\rightarrow \varphi _{I} \) at \( \xi =\mathbf{n}\cdot \mathbf{d}\rightarrow -\infty  \),
\( \varphi _{\mathbf{n}}\rightarrow \varphi _{II} \) at \( \xi \rightarrow \infty  \),
and providing minimum to the energy functional:\begin{equation}
\label{6}
E\left[ \varphi _{\mathbf{n}},\psi _{\mathbf{n}}\right] =\sum _{\mathbf{n}}V_{\mathbf{n}}\left( \varphi _{\mathbf{n}},\psi _{\mathbf{n}}\right) ,\qquad V_{\mathbf{n}}\left( \varphi _{\mathbf{n}},\psi _{\mathbf{n}}\right) =\sum _{\rho }V_{\mathbf{n},\mathbf{n}+\rho }\left( \theta _{\mathbf{n}},\theta _{\mathbf{n}+\rho }\right) .
\end{equation}
Using the above numerical simulation to estimate AIMI, we conclude that the
function \( V_{\mathbf{n}}\left( \varphi _{\mathbf{n}},\psi _{\mathbf{n}}\right)  \)
is well approximated by the sum of symmetric and antisymmetric parts: \( V_{\mathbf{n}}^{\left( s\right) }\left( \psi _{\mathbf{n}}\right) +V_{\mathbf{n}}^{\left( as\right) }\left( \varphi _{\mathbf{n}}\right)  \).
Then the antisymmetric part proves to be the softest mode so that the energy
functional, Eq. \ref{6}, in continuos approximation, \( \varphi _{\mathbf{n}}\rightarrow \varphi \left( \xi \right)  \),
can be written as\begin{equation}
\label{7}
E\left[ \varphi ,\frac{\partial \varphi }{\partial \xi }\right] =\int _{-\infty }^{\infty }\left[ \frac{1}{2}a^{2}v_{1}\left( \frac{\partial \varphi }{\partial \xi }\right) ^{2}+\frac{1}{36}v_{2}\cos 6\varphi \right] .
\end{equation}
In this approximation, the inhomogeneous re-orientations of hexagons across
the domain boundary can be described by the sine-Gordon equation:\begin{equation}
\label{8}
\frac{\partial ^{2}\varphi }{\partial \xi ^{2}}+\frac{1}{3}d_{DW}^{-2}\sin 6\varphi =0,
\end{equation}
where the domain wall width \( d_{DW}=a\sqrt{v_{1}/v_{2}} \) is of the order
of lattice constant \( a \). This is related to the fact that, unlike the common
situation in magnets where as a rule \( v_{1} \) (exchange, or stiffness constant)
is much greater than \( v_{2} \) (relativistic anisotropy), in the considered
system both constants have the same origin in inter-molecular interactions and
hence the same order of magnitude. Though, strictly speaking, Eq. \ref{8} in
this situation is only valid far enough from the domain boundary, orientations
of discrete hexagons (obtained from a certain infinite discrete set of equations)
will follow the {}``kink{}'' solution \( \varphi \left( \xi \right) =\left( 1/3\right) \arcsin \tanh \left( \xi /2d_{DW}\right)  \)
with sufficient accuracy. The factor 1/3 here and in Eq. \ref{8} provides a
correct asymptotics for \( \varphi \left( \xi \right)  \): \( \varphi \left( \pm \infty \right) =\pm \pi /6 \).
This also expresses the analogy of the \( \pi /3 \) rotation between the considered
domains with the \( \pi  \) rotation between 180\( ^{\circ } \) domains in
ferro- and antiferromagnets.

\begin{figure}
{\par\centering \includegraphics{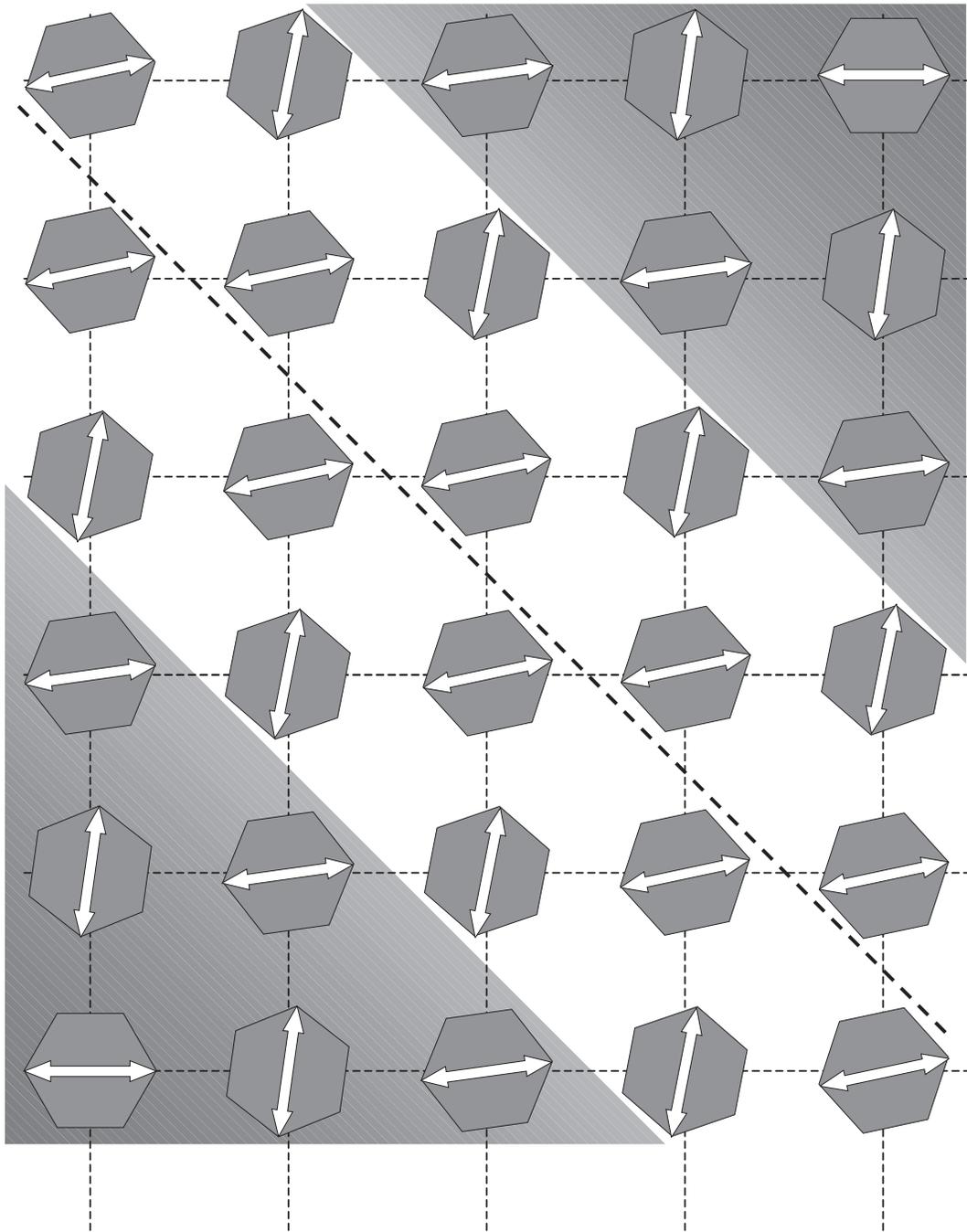} \par}

\caption{\label{fig6}Domain wall (clear region) between two domains (dark regions,
I and II of Fig. \ref{fig5}). For convenience, the directors show the molecules
orientations tilted with respect to the related asymptotics. The dashed line
corresponds to the order parameter \protect\( \varphi =0\protect \).}
\end{figure}

Fig. \ref{fig6} presents an example of a relatively narrow domain wall in the
\emph{sp} lattice of hexagons. A sensible rotation of molecules with respect
to their orientations in domains occurs within the stripe of \( \sim 2\div 3 \)
lattice parameters (\( d_{DW}\sim a \)), hence the misorientations are localized
just in the domain wall. Notice that the 4-sublattice structure admits a richer
systematics of domains (up to 4) and domain boundaries between them.

\begin{figure}
{\par\centering \includegraphics{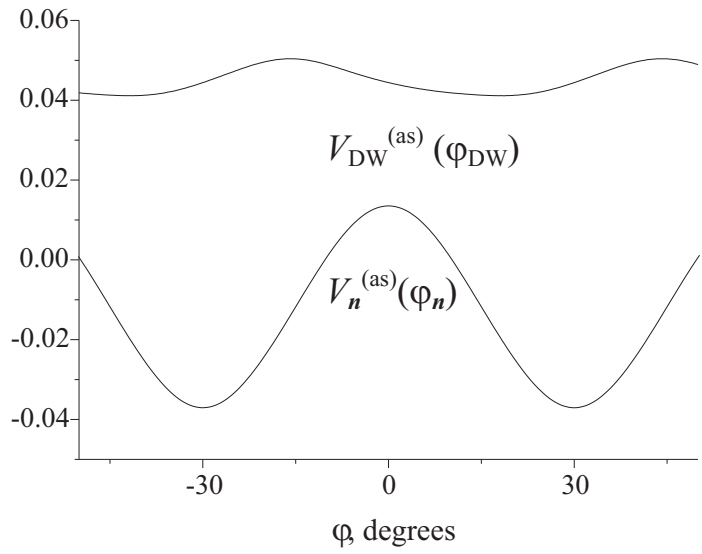} \par}

\caption{\label{fig7}Potential reliefs for the antisymmetric parts of AIMI in a domain
and in the center of domain wall.}
\end{figure}

To examine how the dynamics of misoriented molecules differs from those in the
domain, we estimated the antisymmetric part of AIMI, \( V_{DW}^{\left( as\right) }\left( \varphi _{DW}\right)  \),
for the closest unit cell to the center of domain wall. The corresponding potential
relief shown in Fig. \ref{fig7} is noticeably smoother and its minima, having
the same \( \pi /3 \) periodicity, are much flatter than those for \( V_{\mathbf{n}}^{\left( as\right) }\left( \varphi _{\mathbf{n}}\right)  \).
Therefore the \char`\"{}orientational defects\char`\"{}, or molecules in domain
wall, should display a softer libration spectrum with increasing density to
lower energies. Besides, a specific low-energy excitation mode can appear, corresponding
to oscillations of the antisymmetric order parameter \( \varphi _{DW} \) which
propagate along the wall (a \char`\"{}bending\char`\"{} mode of orientational,
not translational, origin).

The mentioned characteristics of a domain wall can be important for low-temperature
behavior of the crystal. First of all, the collective defects should be stronger
scatterers for thermal phonons than any point defects, especially if the phonon
wavelength (\( \lambda _{T}\sim \hbar v_{s}/(k_{B}T) \), where the sound velocity
\( v_{s}\sim 3\cdot 10^{5} \)cm/s \cite{1}) becomes comparable with \( d_{DW} \).
Besides, a weaker AIMI in domain walls can permit the molecules there to remain
almost free rotators down to much lower temperature than that of orientation
freezing for the rest of the crystal.

\section{Concluding remarks}

The presented consideration shows how peculiar can be dynamics of low-energy
excitations in such a simple model system as that of hexagons on square lattice.
In particular, for asymmetrical hexagons (possessing \( C_{3} \) axis) this
lattice turns out frustrated, which does not exclude a possibility for its glassy
behavior. But even the frustrated lattice can be divided into two sublattices,
leading to domain structure and domain walls. The latter, being of orientation
origin, are able to effectively scatter the excitations of another origins,
in particular, the phonons. However, the detailed analysis of such scattering
goes beyond the scope of this work.

It seems that above results could be also relevant for fullerite. First of all,
this relates to the possibility that \( C_{6} \) rotation around fixed orientation
of each C\( _{60} \) molecule in the \emph{sc} phase could become effectively
an element of the point symmetry group of the averaged crystal. Though the energy
degeneracy conditioned by corresponding random \char`\"{}transverse\char`\"{}
fullerene orientations also admits the existence of orientation glass state
of fullerite, but the crystal as a whole remains uniform and no reasons can
be found for any distinct point defects, including mis-orientations. However
the extended topological defects, like orientation domain walls (which should
not perturb initial attribution of C\( _{60} \) molecules to cube diagonals),
can exist even in a homogeneous system and provide an effective channel for
dissipation of low energy quasi-particles.

Certainly, a more detailed theoretical study of these issues demands more realistic
models, adequate to the fullerene and fullerite structures.

\section{Acknowledgments}

One of us (V.M.L.) gratefully appreciates the suggestions from A.I. Aleksandrovskii,
V.G. Manzhelii, and L.P. Mezhov-Deglin who attracted his attention to the original
experimental results concerning thermal anomalies in fullerites. He also expresses
his gratitude to J. Lopes dos Santos and Centro do Física do Porto (Portugal)
for kind hospitality permitting this work to be done. It was supported in parts
by NATO grant CP/UN/19/C/2000/PO, Program OUTREACH (V.M.L.) and Portuguese project
PRAXIS XXI 2/2.1/FIS/302/94 (Yu.G.P.).

\end{document}